# Machine Learning for Mie-Tronics

Wenhao Li[1], Hooman Barati Sedeh[1], Willie J. Padilla[1], Simiao Ren[1,2], Jordan Malof[2], Natalia M. Litchinitser[1, *]

[1]Department of Electrical and Computer Engineering, Duke University, Durham, NC, USA

[2]Department of Computer Science, University of Montana, Missoula, MT, USA

*natalia.litchinitser@duke.edu

**Abstract**

Electromagnetic multipole expansion theory underpins nanoscale light-matter interactions, particularly within subwavelength meta-atoms, paving the way for diverse and captivating optical phenomena. While conventionally brute force optimization methods, relying on the iterative exploration of various geometries and materials, are employed to obtain the desired multipolar moments, these approaches are computationally demanding and less effective for intricate designs. In this study, we unveil the potential of machine learning for designing dielectric meta-atoms with desired multipolar moments up to the octupole terms. Specifically, we develop forward prediction models to unravel the intricate relationship between the scattering response and the topological attributes of individual meta-atoms, and an inverse design model to reconstruct scatterers with the targeted multipolar moments. Utilizing a tandem network trained to tailor dielectric meta-atoms for generating intended multipolar moments across a broad spectral range, we further demonstrate the generation of uniquely shaped meta-atoms for exciting exclusive higher order magnetic response and establishing super-scattering regime of light-matter interaction. We also illustrate the accurate prediction of electric field distributions within the given scatterer. Our versatile methodology can be readily applied to existing datasets and seamlessly integrated with various network architectures and problem domains, making it a valuable tool for the design of different platforms at nanoscale.

**Introduction**

Rapid progress in photonics and nanofabrication has opened up new prospects for realizing engineered scatterers to manipulate light at the nanoscale [1]. These subwavelength particles can be arranged in isolated, two- and three-dimensional arrangements, known as meta-atoms [2], metasurfaces [3], and metamaterials [4], respectively, and have been shown to facilitate various applications such as beam steering [5], holography [6], nonlinear harmonic generation [7], and Kerker, anti-Kerker, and transverse Kerker effects [8], to name a few. While electromagnetic multipole expansion theory, a cornerstone of light-matter interactions, has facilitated the study of these intriguing optical phenomena, brute force optimization methods relying on an iterative exploration of various geometries and materials, are conventionally employed to obtain the desired response [9-13]. However, these approaches are computationally demanding and less effective for intricate designs, which leads to a fundamental trade-off between performance and time, highlighting the need for alternative methods that offer faster and more efficient solutions to overcome these limitations.

In recent years, machine learning (ML) models have significantly evolved and lead to numerous breakthroughs in various domains such as finance [14], healthcare [15], computer vision [16], and robotics [17]. Followed by such a fruitful progress in this field of research, ML has recently emerged as a powerful tool in photonics for the design and analysis of various subwavelength platforms [18-29]. In particular, contrary to the conventional numerical approaches, ML methods not only offer fast prediction techniques that facilitate the optimization process within high-degree design spaces, but also serve as a valuable technique in tackling inverse design challenges by predicting nanostructure geometries that fulfill specific optical property criteria. Although there has been considerable effort dedicated to implementing ML for designing optical metasurfaces [30-36], the study of light-matter interactions with isolated meta-atoms and their design using ML remains relatively scarce and not yet fully explored. In this context, Wiecha et al. [37] recently demonstrated an example of such an approach, showcasing the feasibility of predicting electromagnetic fields within subwavelength scatterers operating at fixed wavelengths. Along this line of research, there are a few other studies that have employed ML for meta-atom design involving regular and simple geometries such as spheres [38], core-shell structures [39-42], cubes [43], and cylinder [44] which typically involve only the low-order multipolar resonances. However, these works are limited in scope, as designing irregular shapes using ML is more practical and advantageous due to their support of higher-order multipolar moments, which can provide richer physics and enable applications that require high confinement of electromagnetic fields such as enhancing nonlinear conversion efficiency [45] and establishing strong coupling in two-dimensional materials [46]. Furthermore, most of these studies focus on the optical scattering response of meta-atoms without explaining the underlying physical mechanisms and the involving resonant modes.

Here, we harness the power of machine learning to design dielectric meta-atoms supporting desired multipolar moments as it is schematically illustrated in **Figure 1**. In particular, we establish a connection between the induced electromagnetic field, rather than focusing on the optical performance, and the topological features of meta-atoms, via employing multipole expansion theory to expand electromagnetic fields on a multipolar basis [9-13]. By developing forward prediction models (FPM) to decipher the intricate relationships between scattering response and topological attributes of individual meta-atoms (labeled with blue color in Figure 1) and an inverse design model (IDM) to reconstruct scatterers with desired multipolar moments (denoted with red), we demonstrate the effectiveness of our methodology in generating uniquely shaped meta-atoms with tailored functionalities. By training a tandem network for generating intended multipolar moments across a broad spectral range, we demonstrate the creation of uniquely shaped meta-atoms for exciting exclusive higher order magnetic response and establishing super-scattering regime of light-matter interaction. Additionally, we construct a machine learning model to predict the wavelength-dependent electric field distribution of the meta-atom and its surroundings as it is highlighted with pink color in Figure 1. Our versatile methodology can be readily applied to existing datasets and seamlessly integrated with various network architectures and problem domains, making it a valuable tool in nanophotonics.

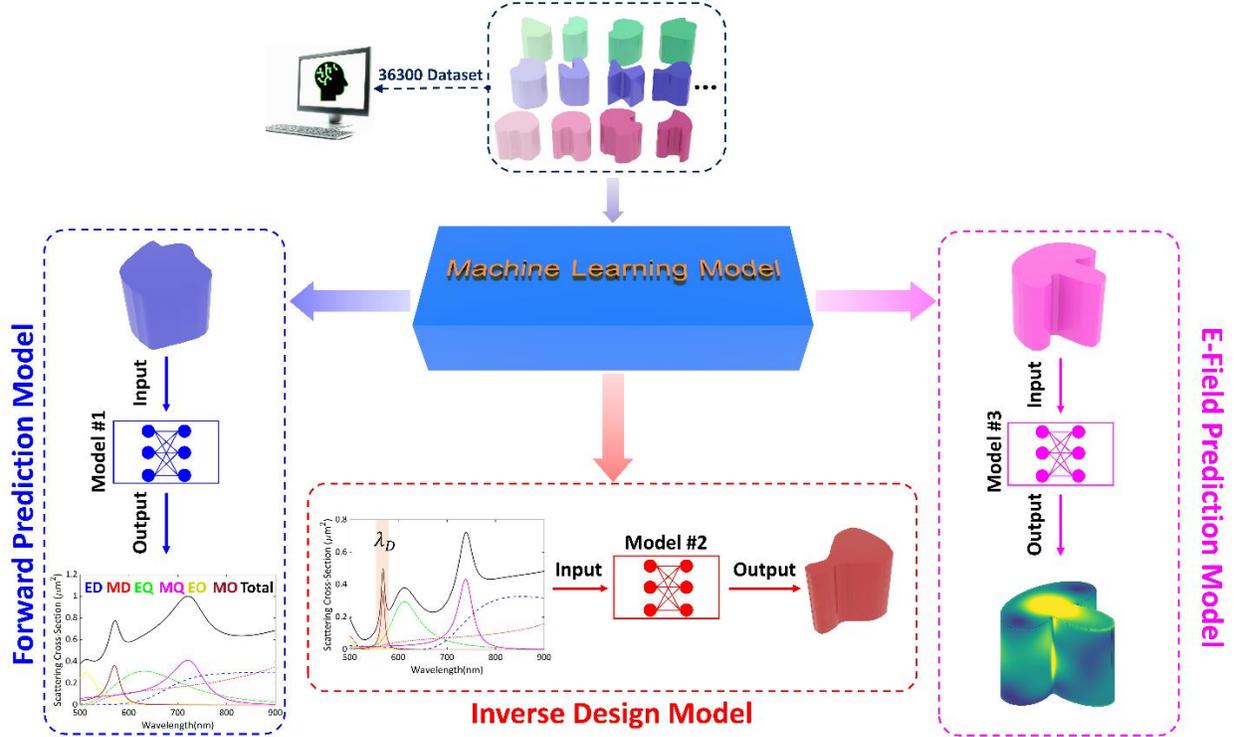

**Figure 1. Artistic illustration of the concept of machine learning for mie-tronics.** The total dataset contains 36300 distinct combinations of shapes and wavelengths in the spectral range of $500 \text{ nm} < \lambda < 900 \text{ nm}$. The height of $TiO_2$ meta-atoms is fixed to $H = 320$ nm while its cross-section contour varies in the transverse plane as function of azimuthal angle. In the FPM scenario, depicted in blue, the meta-atom geometry serves as the input, allowing the model to predict its optical response in terms of multipole moments up to the magnetic octupole. Conversely, during the inverse design phase, indicated by the red color, the model receives the desired multipolar moments at a specific wavelength as input, and accurately predicts the explicit geometry that yields such a response. The third model on the other hand predicts the electric field distribution of the given shape at the desired operating wavelength (shown with pink color curve).

## Results

**Forward Prediction Model.** Upon the interaction of light with an optical scatterer, the induced polarization is related to the field distributions inside the particle through $\boldsymbol{P} = \epsilon_0(\epsilon_p - \epsilon_d)\boldsymbol{E}_p$, where $\epsilon_0$, $\epsilon_p$ and $\epsilon_d$ denote the free space, particle, and surrounding medium dielectric constants, respectively, and $\boldsymbol{E}_p$ represents the total electric field within the meta-atom [9-13]. Since the induced polarization can be described in terms of the induced multipoles, understanding their dynamics offers significant insights into the induced fields inside the nanoresonator. While computing multipole moments using available commercial software can be a time-consuming process, particularly for complex structures, ML offers a more efficient alternative, provided it can be trained to accurately predict the induced resonant modes for a given geometry. For this purpose, we developed an FPM capable of predicting multipolar resonances based on the shape of a meta-atom in the target spectral range of 500 nm to 900 nm. Our developed FPM employs a densely connected convolutional network (DenseNet) encoder architecture [47], which is orders of magnitude faster than conventional commercial software simulations. In this approach, the meta-atom shape information (shape (i)) is compressed into a reduced dimension (bottleneck) following a series of down conversion blocks (DCBs). This condensed shape information, together with the associated wavelength, is then passed through fully connected dense layers to predict the multipolar resonances at a given wavelength, as it is depicted in **Figure 2(a)**.

A dataset of 36300 distinct combinations of titanium dioxide ($TiO_2$) meta-atom shapes and wavelengths are used to train and validate the ML model. The height of meta-atoms is fixed to 320 nm, while their geometrical cross-sections in the transverse plane vary as function of azimuthal angle (see Supplementary Information for more details on the utilized geometries). For calculating the electric field distribution and the induced multipoles within the $TiO_2$ meta-atoms to train the ML model, numerical simulations are carried out using the finite-element method (FEM) implemented in the commercial software COMSOL Multiphysics at 21 discrete points within the spectral window of interest. Contrary to most prior studies, which primarily consider the first four multipole moments to characterize the optical response of regular-shape optical scatterers [37-44], our study extends the consideration to higher-order terms up to the magnetic octupole (see **Methods** for details). This extension is necessitated by the fact that the meta-atom irregular morphology can result in the excitation of higher-order moments, thereby demanding a more comprehensive analysis. Figure 2(b) illustrates the scattering cross-section of a representative $TiO_2$ meta-atom, employed in training the ML model (shape (ii) in panel (a)), across the spectral range of 500 nm to 900 nm. As can be seen, the optical response of the scatterer at various wavelengths is predominantly governed by distinct types of moments. Specifically, the response of the particle can be primarily attributed to the electric and magnetic dipoles at longer wavelengths, while contributions from higher-order moments become increasingly evident with decreasing wavelength. The emergence of higher moments at shorter wavelengths can be ascribed to the particle's relative size in this region being commensurate with the operating wavelength. It is worth mentioning that a comparison between the total scattering response of the meta-atom and the response derived from the summation of the first six multipoles, including electric and magnetic dipoles, quadrupoles, and octupoles (ED, MD, EQ, MQ, EO, and MO, respectively), reveals a high level of accuracy in capturing the optical behavior of the meta-atom. This observation is substantiated by the near-identical nature of the two scattering responses (shown in Figure 2(b)), emphasizing the sufficiency of considering these six multipole moments in describing the meta-atom's optical characteristics accurately and neglecting the multipolar terms higher than MO. For further validation of the multipole resonance assessment, we have also compared the results obtained from the multipole expansion with those derived from Mie theory for light scattering of spherical particles (see Supplementary Information for more details of this comparative analysis). Figure 2(c) demonstrates the average values, their corresponding standard deviations, and types of the resonant modes supported by the utilized training dataset. A noteworthy observation reveals that at longer wavelengths, low order Mie-type resonances, in particular ED and MD moments, exhibit a more pronounced contribution to the overall dataset, whereas at shorter wavelengths higher-order moments such as MO, are emerged. The performance of the predictive model is assessed using the mean square error (MSE) loss metric, which quantifies the deviation between predicted and actual resonance values. In particular, the MSE loss for the predicted resonances in the validation dataset is calculated to be $6.5 \times 10^{-4} \mu m^2$, which is more than two orders of magnitude lower than the average of the resonances, demonstrating the competency of the employed model to generate precise predictions with minimal error relative to the overall resonance values (see Supplementary Information for more details).

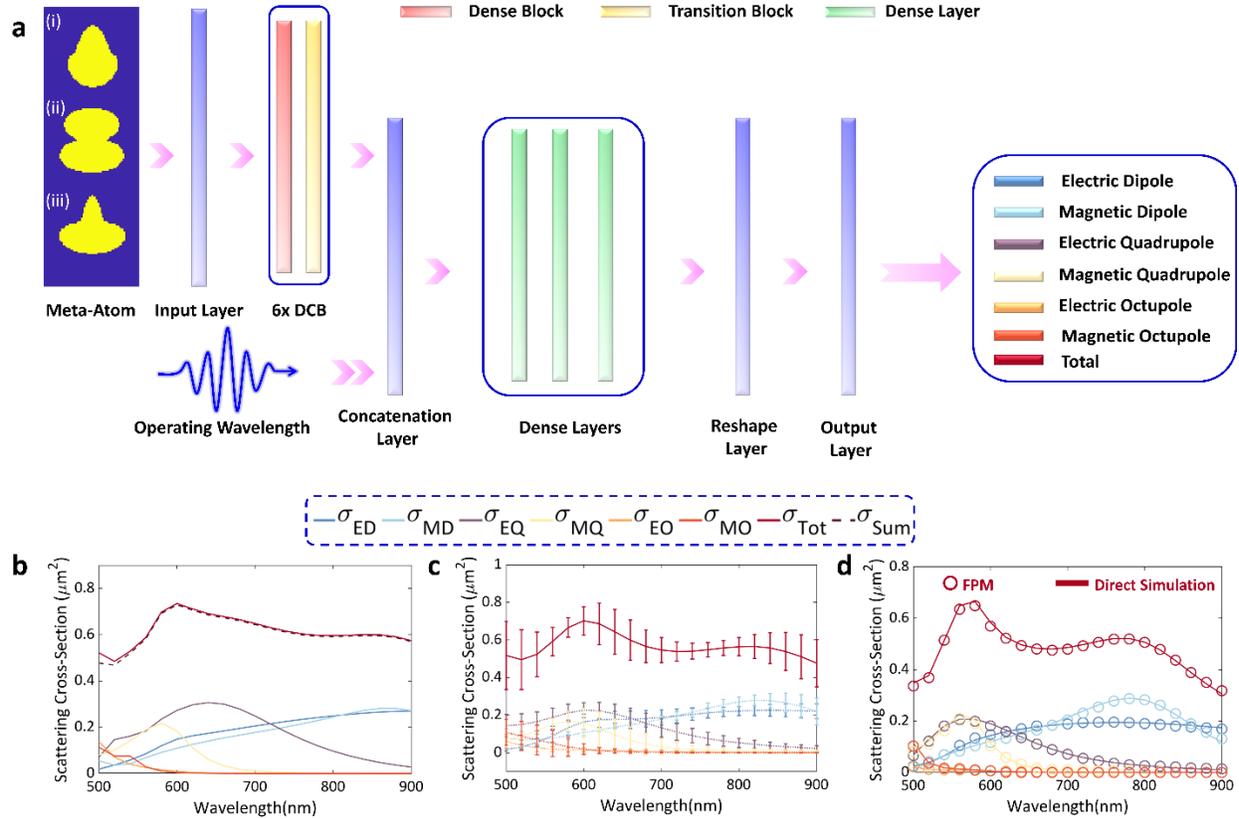

**Figure 2. Predicting meta-atom optical response with a forward prediction model. (a)** The schematic representation of an FPM to predict the optical response of the input meta-atom. The shape information undergoes compression into a reduced dimension via multiple DCBs. Subsequently, this condensed information, in conjunction with the corresponding wavelength, is processed through fully connected dense layers to accurately predict multipolar resonances for a specified wavelength. The scattering cross-section of **(b)** a single meta-atom, second geometry shown in panel (a), and **(c)** the entire training dataset utilized to train the ML model, up to MO resonant mode in the spectral window of 500 nm to 900 nm. **(d)** A comparison of the scattering cross-section between direct simulation results (dashed lines) and predictions from the FPM (solid line) for an arbitrary meta-atom given as the input meta-atom (third geometry in panel (a)).

Figure 2(d) presents a comparison of the resonances of a meta-atom (marked by (iii) in panel (a)) obtained through direct simulation (solid line) and those predicted by FPM (dashed line) across 21 wavelengths in the spectral range of 500-900 nm. The close agreement between the simulation and FPM-predicted results highlights the remarkable capability of the implemented model to accurately predict the optical response of meta-atoms, which can be used for various applications including advanced metamaterials, photonic devices, and nanoscale optical systems.

**Inverse Design Model.** Despite the efficiency and accuracy of our forward prediction model in estimating the induced multipolar resonances of a given meta-atom, addressing the inverse problem— acquiring a geometry corresponding to a desired electromagnetic response—continues to pose a formidable and resource-intensive challenge in the field of nanophotonics. On the other hand, previously reported inverse design models (IDM) are often deemed ill-posed for two main reasons of i) existence - a meta-atom with the given multipolar resonances might not exist, and ii) uniqueness - multiple meta-atoms could exhibit the same multipolar resonances [18-36]. Apart from these challenges, existing IDMs require spectral information across a broad range, which complicates their practical application, as spectral information outside the target range might be unknown yet still required for the input. As a result, adopting the previously reported models for the inverse design of meta-atoms based on their optical responses would intensify the aforementioned issues, given that all resonance curves must be supplied. To address these challenges, we have developed an IDM that streamlines the meta-atom design process and enhances the likelihood of achieving the desired multipolar resonances. In particular, the utilized model employs the desired resonant modes at a specific operating wavelength as the input for predicting the meta-atom geometries, circumventing the problem of requiring spectral information across a broad range. This approach is particularly useful in the context of the ill-posed nature of inverse design problems, as it reduces the complexity associated with the existence challenges.

The uniform manifold approximation and projection (UMAP) of the sample distribution in high-dimensional resonance space is illustrated in **Figure 3(a)** which unveils a discernible pattern resulting from resonance interdependence and significant variation across different wavelengths. We note that the presence of sample clusters in resonance space implies that multiple samples possess closely situated resonances, thus complicating the inverse design process. To overcome such an issue and avoid generating inaccurate designs due to sample non-uniqueness, we implemented a Tandem inverse design model (TIDM), as it is demonstrated in Figure 3(b). In particular, the employed model consists of two components: a DenseNet decoder for designing meta-atom shapes and a pre-trained FPM for validating the resonances of the meta-atom design. To ensure the design convergence, the MSE loss between the desired resonances (i.e., target response) and the ones obtained from FPM is implemented as the feedback for updating the decoder weights while maintaining its FPM counterparts fixed. Within this configuration, the IDM is compelled to select a single realization of meta-atoms capable of generating the desired resonances while ignoring alternative possibilities.

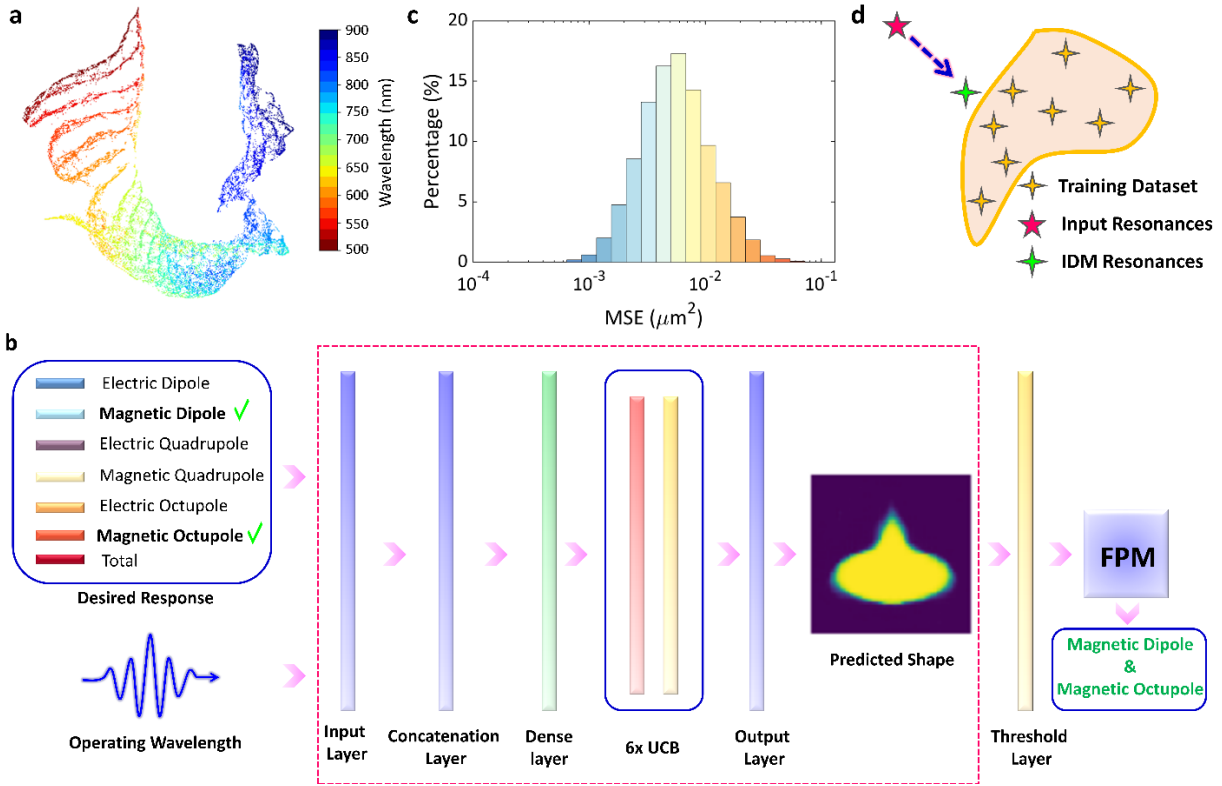

**Figure 3. Meta-atom implementation using inverse design model. (a)** A UMAP of the sample distribution in high-dimensional resonance space, illustrating the interdependence of resonances and their variation across a spectrum of wavelengths. **(b)** The schematic diagram showcasing the implemented IDM used for designing meta-atom geometry based on the desired optical response at a specified operating wavelength. The MSE loss between the target response and those predicted by the FPM is incorporated as feedback to ensure the identification of an optimal solution. **(c)** A histogram depicting the MSE between the predicted and designed resonances, demonstrating an average error of $0.007~\mu m^2$. **(d)** Conceptual visualization of the strategic approach undertaken to address potential non-existence problems that may arise during the inverse design process. Upon the employed method, the designed meta-atoms approach the training data pool.

To evaluate the performance of the inverse design model, meta-atoms are designed using the wavelength and resonances from the validation dataset, and their resonances are predicted using FPM. Figure 3(c) demonstrates the histogram of the MSE between the predicted and designed resonances, with an average error of $0.007~\mu m^2$, which is two orders of magnitude lower than the average of the resonances. It should be remarked that in the process of designing a meta-atom, it is inherently uncertain whether meta-atoms with specified resonances exist or not. Moreover, in the event that such meta-atoms could be found, their characteristics could diverge considerably from those of the training samples, leading to an underwhelming performance of the IDM. To overcome such an issue, we implemented a method such that when the IDM is provided with resonances that significantly deviate from those in the training dataset, it tends to generate results that align closely with the training dataset examples within a specific confidence range as it is shown in Figure 3(d) (see Supplementary Information for more details). To achieve the optimal design, we explore the parameter space and evaluate the device performance using FPM. In practice, all weights in the tandem IDM remain fixed, while the input parameters are updated through backpropagation. To demonstrate the efficacy of the presented IDM, we engineer various scatterers hosting strong magnetic multipolar responses and super-scattering as representative examples.

**Magnetic octupolar response.** The realization of efficient optical magnetism using nonmagnetic nanostructures is crucial for controlling light-matter interactions at the nanoscale. This capability has advanced numerous significant research fields, including Fano resonances [48-50], enhanced optical nonlinearity [51], directional scattering of light [52-54], and Purcell factor enhancements [55] among others. In this perspective and considering the pivotal role of optical magnetism in nanophotonics, it is essential to design a scatterer wherein the magnetic response, such as MD, MQ, and/or MO, constitutes the dominant contribution relative to other resonant modes. Although to date, various methods have been proposed to attain this particular regime of light-matter interaction, these approaches typically involve constraints in terms of geometry, such as core-shell structures [56], or depend on the specific excitation of non-radiating states, such as optical anapoles [57]. Here we demonstrate the capability of our IDM to predict and design two different scatterers supporting magnetic responses (MD and MO) at two distinct operating wavelengths. For this purpose, the loss functions for each of the representing examples are set to be $L_{\text{MD}} = -\sigma_{\text{MD}}/\sigma_{\text{Tot}}$ and $L_{\text{MO}} = -\sigma_{\text{MO}}/\sigma_{\text{Tot}}$. These loss functions are subsequently employed to iteratively update the resonant characteristics within the input layer until they converge to a minimum value. Once this convergence is achieved, the design of the meta-atom is extracted from the optimized parameters. Employing the outlined method, we observed that the MD resonant mode's contribution for the first scatterer, at an operating wavelength $\lambda_{\text{MD}} = 780$ nm, accounts for nearly 61% of the total response as it is shown in **Figure 4(a)**. We note that at the corresponding wavelength of $\lambda_{\text{MD}}$, the contribution of MD is almost two times higher than that of the ED moment, i.e., $\sigma_{\text{MD}}/\sigma_{\text{ED}} \approx 2$. Similarly, the second meta-atom, designed to support MO at $\lambda_{\text{MO}} = 500$ nm, contributes approximately 50% of the total scattering cross sections in the optimized design, which represents a higher contribution than the best currently available training dataset as it is shown in Figure 4(b) (see Supplementary Information for more details on the optical response of the MO scatterer in the region of $450 \text{ nm} < \lambda < 550 \text{ nm}$). We note that to the best of our knowledge, achieving such a high degree of magnetic contribution, particularly for higher order moments such as MO, has not been documented in any prior research. In addition to the scattering cross-section spectra, we have depicted the near field distribution of the scatterers at their respective MD and MO resonant modes' operating wavelengths in Figures 4(c) and 4(d), respectively. We note that the observed deviations between the predicted response and the ideal distribution of MD and MO modes can be ascribed to the concurrent presence of other states at the identical spectral position as that of the desired magnetic responses.

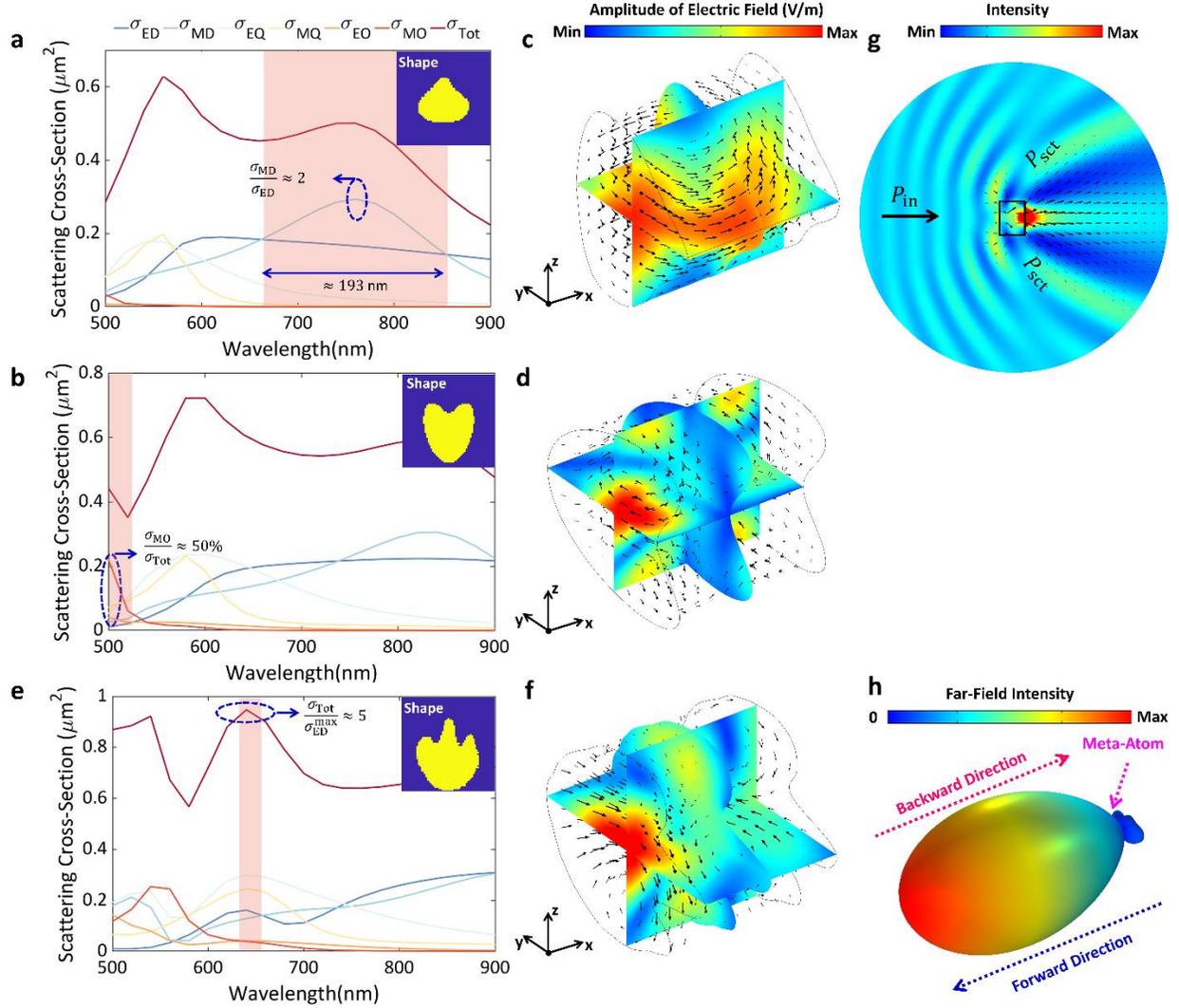

**Figure 4. Engineering magnetic and super-scattering responses through IDM.** The scattering spectra of the designed meta-atoms for supporting **(a)** MD and **(b)** MO with the contribution of 61% and 50% of the total response at the operating wavelengths of $\lambda_{MD} = 780$ nm and $\lambda_{MO} = 500$ nm, respectively. The three-dimensional near-field distribution of the representative examples supporting **(c)** MD and **(d)** MO resonant modes, with black arrows representing the electric field components. **(e)** The scattering spectra of the third meta-atom, designed to support super-scattering state at $\lambda_{SS} = 633$ nm. At the corresponding wavelength, five-fold enhancement of scattering, compared to single-channel limit ($\frac{\sigma_{Total}}{\sigma_{ED}^{max}} \approx 5$) can be observed. **(f)** The amplitude of the electric field within the super-scatterer at $\lambda_{SS} = 633$ nm and **(g)** its corresponding intensity distribution in the background environment. The black arrows in panel (g) represent the relative Poynting vectors, which tend to scatterer in the forward direction. **(h)** The far-field radiation pattern of the super-scatterer at $\lambda_{SS}$. The designed meta-atom significantly scatters in the direction of the incident beam with the maximal directivity $\approx 9$ dB.

**Super-scattering optical response.** Achieving enhanced wave scattering from subwavelength particles is of utmost importance in various applications, including sensors, miniaturized antennas, and energy-harvesting devices [58-63]. However, the successful implementation of these enhancements presents a significant challenge attributed to the inherent limitation associated with subwavelength scatterers, known as single-channel limit [59]. In particular, this constraint establishes an upper boundary for the scattering cross-section of such scatterers which is bounded to $(2l + 1)\lambda^2/2\pi$, wherein $l$ represents the order of the multipole and can only be achieved under resonance conditions for a specific scattering mode or channel [64], [65]. Overcoming this limit requires resonant scattering across multiple modes at a single frequency, resulting in what so-called super-scattering state, which can enhance wave scattering from subwavelength objects. Conventionally, the notation of super-scattering applies to the cases where the total scattering cross-section surpasses the given single-channel limit of an electric dipole ($l = 1$) such that $\frac{\sigma_{\text{Tot}}}{\sigma_{\text{ED}}^{\max}} = \frac{2\pi \sigma_{\text{Total}}}{3\lambda^2} > 1$ [58-65]. Although the initial proposal for achieving a state of super-scattering concentrated on cylindrical structures comprising multiple plasmonic and dielectric layers [59], the experimental results revealed that the measured enhancement in wave scattering was approximately 50% less than predicted (from 8 times enhancement to 4 times) or, in some cases, even non-existent [66], which is attributed to the losses in the plasmonic materials. A similar phenomenon was observed in a core-shell plasmonic nanowire, where the scattering cross section slightly exceeded the single-channel limit in the presence of material losses [66] and both in the terahertz and microwave frequency regimes [67], [68]. To surmount this challenge, we employ our developed IDM to conceive an all-dielectric meta-atom that manifests a state of super-scattering. Remarkably, our proposed structure surpasses the previously reported enhancement values while maintaining a simplified design devoid of multilayers or the utilization of metasurfaces around the scatterer [69]. In contrast to previous studies that aimed to spectrally align the peaks of contributing moments at a single frequency, our approach focuses on directly maximizing the total scattering cross-section (i.e., $\sigma_{Tot}$). This offers several advantages: firstly, it incorporates all potential contributions beyond the MO moment, allowing for a comprehensive analysis. Additionally, it simplifies the process since searching for a geometry that aligns all peaks at a single point can be challenging and may not be feasible. To this end, we defined the corresponding loss function to be $L_{\text{SS}} = -\sigma_{Tot}$, and then it is used to update the resonant characteristics within the input layer until the design of the meta-atom is extracted from the optimized parameters. Within this process, the desired geometry was extracted as it is shown in the inset of Figure 4(e) which possess a total scattering cross section of $\sigma_{\text{Tot}} = 0.94 \ \mu m^2$ (0.945 $\mu m^2$ predicted by FPM) at the operating wavelength of $\lambda_{\text{SS}} = 633$ nm. According to the obtained result, such a scatterer operates in the super-scattering regime with the total enhancement of $\frac{\sigma_{\text{Total}}}{\sigma_{\text{ED}}^{\max}} \approx 5$, which to the best of our knowledge it the highest reported value calculated within a simple all-dielectric platform. We note that that the observed enhancement in scattering is not reliant on the alignment of peaks from individual contributing moments, but rather on their collective overlap as it is shown in Figure 4(e). This significant improvement in scattering arises from the maximization of the total scattering cross-section, which considers the overall contribution of each moment, rather than focusing solely on individual moments. Figures 4(f) and 4(g) show the near-field and intensity distributions of the scatterer at the corresponding wavelength of $\lambda_{\text{SS}}$ within the meta-atom and its background environment, respectively. The directional scattering behavior of the designed meta-atom is clearly evident from the components of the relative Poynting vector in panel (g). Specifically, the meta-atom exhibits a strong tendency to scatter light in the forward direction, while minimizing scattering in the backward direction. This unidirectional scattering behavior is also reflected in the meta-atom's radiation pattern, which exhibits a high degree of directivity, as it is shown in Figure 4(h). As can be seen from this panel, with a maximal directivity of 9 dB, the majority of the scattered light is concentrated in the forward direction, while the scattering in the backward direction is effectively suppressed. We note that the directivity of the meta-atoms has been calculated using $D = \frac{4\pi\, U(\theta,\varphi)}{\int_0^{2\pi}\int_0^{\pi} U(\theta,\varphi)\sin(\theta)d\theta d\varphi}$, wherein $U(\theta,\varphi)$ represent the radiation intensity, while $\theta$ and $\varphi$ denote the elevation and azimuthal angles, respectively [70].

**E-Field Prediction Model.** While the FPM and IDM have been valuable in designing meta-atoms for/from far-field scattering spectra, the complete description of meta-atom scattering behavior goes beyond resonance intensities. Understanding the electric field distribution within meta-atoms is essential as it directly influences their scattering characteristics and is a complement to multipolar moments for applications wherein spatial overlap integral of the modes are crucial such as in nonlinear photonics [71]. In nanostructured resonant systems, the efficacy of specific nonlinear processes, such as second or third harmonic generation, is contingent upon the extent of spatial field overlap between the pump field and the field associated with the particular nonlinear process within the regions occupied by nonlinear sources. Hence, by designing scatterers that can facilitate desired field distributions at specific operating wavelengths, one can anticipate maximizing the spatial overlap, consequently resulting in enhanced efficiency for the desired nonlinear process. To mitigate the computational burden associated with simulating the electric field distribution and optimizing the desired geometry, here we have devised a three-dimensional (3D) electric field prediction model (EPM) capable of forecasting the electric field within the given meta-atoms as it is shown in **Figure 5(a)**. In particular, the implemented model employs a 3D DenseNet architecture that takes the shape of the meta-atom and compresses it into reduced dimension using six DCBs. The condensed information, along with the desired operating wavelength, is then passed through three dense layers and six up conversion blocks (UCBs) to accurately predict the electric fields within the given scatterer. More specifically, the output layer consists of six channels, each corresponding to the real and imaginary parts of the electric field components or equivalently the electric field amplitude at each plane.

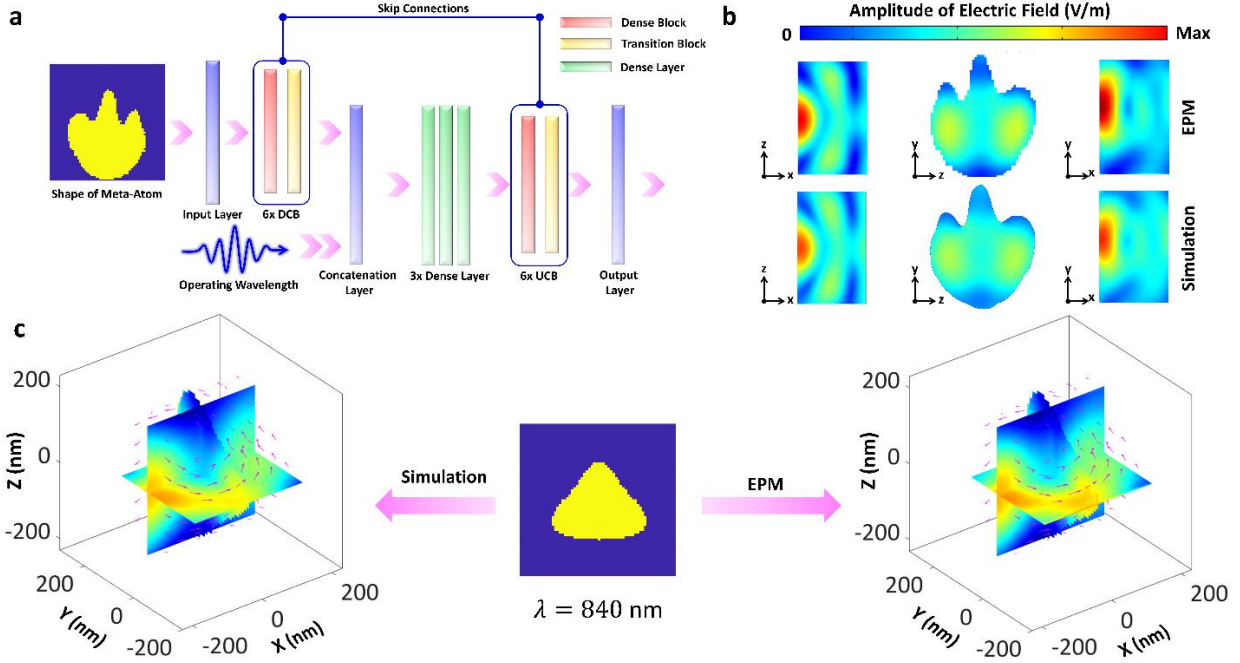

**Figure 5. Electric field prediction model for near-field estimation. (a)** The schematic illustration of the EPM for predicting the near-field distribution of the input meta-atom. The shape information is compressed into a lower dimension using multiple DCBs. The condensed information, along with the corresponding wavelength, is then passed through six dense layers and multiple UCBs to precisely estimate the electric fields within the given scatterer. **(b)** The output of the EPM (top row) and direct simulations (bottom row) for the presented case of super-scatterer in the (x-z), (y-x), and (z-y) planes at the operating wavelength of $\lambda_{SS} = 633$ nm. The same colorbar range is utilized for both cases to ensure consistent visualization and comparison between the two methods (white dashed lines are used for showing the meta-atom boundary). **(c)** A comparison between the 3D near-field distribution of another representative example (middle part) obtained from direct simulation and EPM (left and right panels, respectively). The arrows represent the components of the electric field.

To validate the reliability of the implemented EPM, we utilize the meta-atom geometry of the representative super-scatterer (Figure 4(e)) as input for our model and calculate the electric field amplitude in three distinct planes, namely (x-z), (y-x), and (z-y), at the operating wavelength of $\lambda_{SS} = 633$ nm, as depicted in Figure 5(b) (top row) and compare its results with those acquired through direct simulations (bottom row). As can be seen in this panel, the electric field distributions obtained from the EPM, closely resemble the corresponding distributions from the direct simulations, with the strong agreement between their field values at various positions within the scatterer. To further assess the validity of our developed EPM, we have examined another meta-atom with a distinct geometry, as illustrated in Figure 5(c) (middle part). In this evaluation, we compare the 3D near-field distributions obtained from direct simulations (left side of the panel) with those predicted by the EPM (right part of the panel). Remarkably, the results exhibit excellent agreement between the two approaches, making the EPM a powerful approach to design scatterers with desired field distribution. Additionally, the components of the electric field within the scatterer are provided in the figure, enabling a comprehensive analysis of the field characteristics within the meta-atom (see Supplementary Movie 1 and Movie 2 for additional time-dependent visualizations and analyses of the studied phenomena).

**Discussion**

In summary, this study has successfully utilized machine learning techniques to design dielectric meta-atoms with targeted multipolar moments. By employing the FPM, IDM, and EPM, we achieved accurate prediction, reconstruction, and estimation of the desired scattering responses and near-field distributions. Specifically, the FPM was successfully implemented to provide accurate and efficient predictions of the scattering response for scatterers of arbitrary shapes, eliminating the need for computational software, with high prediction accuracy as indicated by the MSE value of $6.5 \times 10^{-4} \ \mu m^2$. Moreover, the IDM was employed to design three representative scatterers, showcasing their ability to support magnetic octupolar response and achieve super-scattering states. In particular, it was shown that the scatterers exhibited significant contributions of MD and MO moments at two distinct operating wavelengths, with their combined response reaching up to 61% and 50% of the total scattering response, respectively. Additionally, the designed super-scatterer demonstrated nearly five-fold enhancement of scattering beyond the single-channel limit. The EPM was shown to accurately predict the field distribution within the given meta-atoms, allowing for the maximization of spatial mode overlap integral, which is particularly relevant in applications such as nonlinear photonics. **Table 1** provides a comprehensive comparison between the method proposed in this paper and other existing works, taking into consideration various aspects such as materials, shapes, and the underlying physics involved in light-matter interaction. The presence of a cross symbol in Table 1 indicates the absence of a specific response or characteristic in the corresponding work. This comparison serves to highlight the distinct features and advantages of the method presented in this study in relation to other approaches in the field. The versatile methodology presented in this study offers wide applicability, as it can be easily applied to existing datasets and seamlessly integrated with diverse network architectures and problem domains. This adaptability makes it a valuable tool for designing various nanoscale platforms, paving the way to explore different applications and optimize performance in a range of contexts.

Table 1. A comparative analysis of the presented study against existing literature"

| Reference | Materials | Structure | Physics | FPM | IDM | EPM |
|---|---|---|---|---|---|---|
| [39-42] | Plasmonic-Dielectric | Core-Shell | Only $\sigma_{Tot}$ | ✗ | ✓ | ✗ |
| [44] | Plasmonic-Dielectric | Multilayer Cylinder | Up to $\sigma_{MQ}$ | ✗ | ✓ | ✗ |
| [37] | Plasmonic and Dielectric | Arbitrary Shape | Only $\sigma_{Tot}$ | ✓ | ✗ | ✓ |
| [38] | Dielectric | Sphere | Only $\sigma_{Tot}$ | ✗ | ✓ | ✗ |
| [43] | Plasmonic and Dielectric | Cube | Only $\sigma_{Tot}$ | ✓ | ✗ | ✗ |
| This work | Dielectric | Arbitrary Shape | Up to $\sigma_{MO}$ | ✓ | ✓ | ✓ |

## Methods

**Numerical Simulations:** The numerical simulations are carried out using the finite-element method (FEM) implemented in the commercial software COMSOL Multiphysics. In particular, we utilize the Wave Optics Module to solve Maxwell's equations in the frequency domain together with proper boundary conditions. Here, we use a spherical domain filled with air and a radius of $4\lambda$ as the background medium, while perfectly matched layers of thickness $0.6\lambda$ are positioned outside of the background medium to act as absorbers and avoid back-scattering. Tetrahedral mesh is also chosen to ensure the accuracy of the results and allow numerical convergence. The meta-atoms are under plane wave excitation propagating along $x$-axis with its electric field pointing toward $y$-axis.

**Multipole Decomposition:** According to the multipole expansion theory, upon the interaction of light with subwavelength particles, the scattered field in the far-field can be expressed as the superposition of various multipoles (up to the electric octupole term) as [9-13]

$$\boldsymbol{E}_{sct}(\boldsymbol{n}) = \frac{k_0^2 \exp(ik_0 r)}{4\pi\epsilon_0 r} \Bigg( [\boldsymbol{n} \times [\boldsymbol{D} \times \boldsymbol{n}]] + \frac{1}{c}[\boldsymbol{m} \times \boldsymbol{n}] + \frac{ik_0}{6}[\boldsymbol{n} \times [\boldsymbol{n} \times \hat{Q}\boldsymbol{n}]] + \frac{ik_0}{2c}[\boldsymbol{n} \times \widehat{M}\boldsymbol{n}] \\ + \frac{k_0^2}{6}[\boldsymbol{n} \times [\boldsymbol{n} \times \hat{O}^{(e)}(\boldsymbol{nn})]] + \frac{k_0^2}{6c}[\boldsymbol{n} \times [\boldsymbol{n} \times \hat{O}^{(m)}(\boldsymbol{nn})]] \Bigg), \quad (1)$$

where $\boldsymbol{D}$ corresponds to the exact total electric dipole (TED), $\boldsymbol{m}$ is the exact magnetic dipole (MD) moment, and $\hat{Q}$, $\hat{O}^{(e)}$, $\widehat{M}$ and $\hat{O}^{(m)}$ represent the electric quadrupole (EQ), electric octupole (EO), magnetic quadrupole (MQ) and magnetic octupole (MO) tensors, respectively; $\boldsymbol{n} = \boldsymbol{r}/r$ is the unit vector directed from the particle's center towards an observation point, and $c$ and $k_0$ are the speed of light and wavenumber in vacuum. Using these notations, the scattering cross-section, can be written as follows

$$\sigma_{\text{Sct}} \approx \frac{k_0^4}{12\pi\epsilon_0^2 \eta_0 I_0}|\boldsymbol{D}|^2 + \frac{k_0^4 \mu_0}{12\pi\epsilon_0 \eta_0 I_0}|\boldsymbol{m}|^2 + \frac{k_0^6}{1440\pi\epsilon_0^2 \eta_0 I_0}\sum_{x_1,x_2}|Q_{x_1 x_2}|^2 + \frac{k_0^6 \mu_0}{160\pi\epsilon_0 \eta_0 I_0}\sum_{x_1,x_2}|M_{x_1 x_2}|^2 \\ + \frac{k_0^8}{3780\pi\epsilon_0^2 \eta_0 I_0}\sum_{x_1,x_2,x_3}|O^{(e)}_{x_1 x_2 x_3}|^2 + + \frac{k_0^8 \mu_0}{3780\pi\epsilon_0 \eta_0 I_0}\sum_{x_1,x_2,x_3}|O^{(m)}_{x_1 x_2 x_3}|^2 \quad (2)$$

where $I_0$ corresponds to the maximum beam intensity in a focal plane, $\eta_0$, $\epsilon_0$ and $\mu_0$ are the impedance, permittivity and permeability of free space, respectively, and $x_1, x_2$, and $x_3$ represent the different components of each tensor. Each of the presented moments in Equation (2) can be expressed in terms of the induced current within the particle ($\boldsymbol{J} = \partial\boldsymbol{P}/\partial t$) as

$$\boldsymbol{D} = \frac{i}{\omega}\int j_0(k_0 r')\boldsymbol{J}(\boldsymbol{r}')d\boldsymbol{r}' + \frac{ik_d^2}{2\omega}\int \frac{j_2(k_d r')}{(k_d r')^2}[3(\boldsymbol{r}' \cdot \boldsymbol{J})\boldsymbol{r}' - r'^2 \boldsymbol{J}]d\boldsymbol{r}',$$

$$\boldsymbol{m} = \frac{3}{2}\int \frac{j_1(k_d r')}{k_d r'}[\boldsymbol{r}' \times \boldsymbol{J}]\,d\boldsymbol{r}', \quad (3)$$

$$\widehat{M} = 5\int \frac{j_2(k_d r')}{(k_d r')^2}([\boldsymbol{r}' \times \boldsymbol{J}] \otimes \boldsymbol{r}' + \boldsymbol{r}' \otimes [\boldsymbol{r}' \times \boldsymbol{J}])d\boldsymbol{r}',$$

$$\hat{Q} = \frac{3i}{\omega} \int \frac{j_1(k_d r')}{k_d r'} \big[3(\mathbf{r'} \otimes \mathbf{J} + \mathbf{J} \otimes \mathbf{r'}) - 2(\mathbf{r'} \cdot \mathbf{J})\bar{\bar{I}}\big] d\mathbf{r'}$$
$$+ \frac{i6k_d^2}{\omega} \int \frac{j_3(k_d r')}{(k_d r')^3} \big[5(\mathbf{r'} \cdot \mathbf{J})\mathbf{r'} \otimes \mathbf{r'} - r'^2(\mathbf{J} \otimes \mathbf{r'} + \mathbf{r'} \otimes \mathbf{J}) - (\mathbf{J} \cdot \mathbf{r'})r'^2 \bar{\bar{I}}\big] d\mathbf{r'},$$

$$\hat{O}^{(e)} = \frac{15i}{\omega} \int \frac{j_2(k_d r')}{(k_d r')^2} \big(\mathbf{J} \otimes \mathbf{r'} \otimes \mathbf{r'} + \mathbf{r'} \otimes \mathbf{J} \otimes \mathbf{r'} + \mathbf{r'} \otimes \mathbf{r'} \otimes \mathbf{J} - \hat{A}\big) d\mathbf{r'},$$

$$\hat{O}^{(m)} = \frac{105}{4} \int \frac{j_3(k_d r')}{(k_d r')^3} \big([\mathbf{r'} \times \mathbf{J}] \otimes \mathbf{r'} \otimes \mathbf{r'} + \mathbf{r'} \otimes [\mathbf{r'} \times \mathbf{J}] \otimes \mathbf{r'} + \mathbf{r'} \otimes \mathbf{r'} \otimes [\mathbf{r'} \times \mathbf{J}] - \hat{B}\big) d\mathbf{r'},$$

where $j_l(x)$ is the $l$th order spherical Bessel function, $k_d$ is the wave number in the surrounding medium, $\bar{\bar{I}}$ is the $3 \times 3$ unit tensor, and the operators of $\cdot$ , $\times$, and $\otimes$ represent the scalar, vector, and tensor products, respectively. It should be noted that $\hat{A}$ and $\hat{B}$ are auxiliary tensors whose components are obtained according to $A_{x_1 x_2 x_3} = \delta_{x_1 x_2} V_{x_3} + \delta_{x_1 x_3} V_{x_2} + \delta_{x_2 x_3} V_{x_1}$, and $B_{x_1 x_2 x_3} = \delta_{x_1 x_2} V'_{x_3} + \delta_{x_1 x_3} V'_{x_2} + \delta_{x_2 x_3} V'_{x_1}$ which $x_1 = (x, y, z)$, $x_2 = (x, y, z)$, $x_3 = (x, y, z)$ and $\delta$ is the Dirac delta while $V = 0.2[2(\mathbf{r'} \cdot \mathbf{J}) \otimes \mathbf{r'} + r'^2 \mathbf{J}]$ and $V' = 0.2[\mathbf{r'} \times \mathbf{J}]r'^2$.

**Machine Learning Model:** Forward prediction model (FPM) using Dense Convolutional Network (DenseNet) architecture is developed to predict the multipolar resonances of specific meta-atoms as functions of wavelength [47]. In particular, the meta-atom shapes are input into six down-conversion dense blocks before reaching the bottleneck layer. The FPM output comprises multipolar resonance arrays connected to the bottleneck layer through three consecutive fully connected layers. The advantage of employing DenseNet architecture lies in the fact that each layer within the model can receive inputs from all previous layers while forwarding its feature-maps to all subsequent layers. This machine learning process takes mere seconds, offering a significant speed advantage compared to FEM simulations.

**Acknowledgments**

This paper was supported in part by the Office of Naval Research (ONR) (Grant No. N00014-20-1-2558), National Science Foundation (NSF) (Grant No. 1809518), and Army Research Office Award (Grant No. W911NF1810348).

**Conflict of Interest**

The authors declare no conflict of interest.

**Data Availability Statement**

The data that support the findings of this study are available from the corresponding author upon reasonable request.

**Author contributions**

N.M.L. and W.L. initiated the idea of this study. W.L. and H.B.S. conducted theoretical and numerical studies. W.P.J., S.R., and J.M. contributed to modeling the machine learning design and provided valuable insights during discussions. N.M.L. supervised the study performed in this work. All authors collectively contributed to the writing of the manuscript.